\documentclass[twocolumn,showpacs,preprintnumbers]{revtex4}%
\usepackage{amsfonts}
\usepackage{amsmath}
\usepackage{graphicx}
\usepackage{dcolumn}
\usepackage{bm}
\usepackage{amssymb,epsfig}
\usepackage{amssymb}%
\begin{document}
\title{Deviation from the exponential decay law in relativistic quantum field theory:
the example of strongly decaying particles}
\author{Francesco Giacosa$^{\text{a}}$, and Giuseppe Pagliara$^{\text{b,c}}$}
\affiliation{$^{\text{a}}$Institut f\"{u}r Theoretische Physik, Johann Wolfgang Goethe
Universit\"{a}t, Max-von-Laue-Str.\ 1, D--60438 Frankfurt am Main, Germany}
\affiliation{$^{\text{b}}$Institut f\"{u}r Theoretische Physik,
Ruprecht-Karls-Universit\"at, Philosophenweg 16, D-69120, Heidelberg, Germany}

\affiliation{$^{\text{c}}$Dip.~di Fisica dell'Universit\`a di Ferrara and INFN
Sez.~di Ferrara, Via Saragat 1, I-44100 Ferrara, Italy }

\begin{abstract}
We show that a short-time regime, in which a deviation from the exponential
decay law occurs, exists also in the framework of a superrenormalizable
relativistic quantum field theory. This, in turn, implies the possibility of a
quantum Zeno effect also for elementary decays. The attention is then focused
on the typical order of magnitude of strong decay rates of mesons: for these
particles, strong deviations from the exponential decay law are present during
a period of time comparable with their mean life time. As a concrete example,
the case of the $\rho$ meson is studied.

\end{abstract}

\pacs{03.70.+k, 03.65.Xp, 14.40.Be}
\maketitle

Since the discovery of radioactivity, the theoretical and experimental
understanding of unstable particles has attracted much attention of
physicists. The development of quantum mechanics in the beginning of the 20th
century has been a crucial step in this direction. The decay is, in fact,
subject to a fundamental quantum indetermination: only the probability that it
happened or not can be calculated.

When an unstable state $\left\vert S\right\rangle $ is created/prepared at
$t=0$, the survival probability $p(t)$ is defined as the probability that the
state has not decayed yet at the instant $t$. The usual, empirical exponential
law of decay $p(t)=e^{-\Gamma t}=e^{-t/\tau},$ where $\Gamma$ is the decay
width and $\tau=\Gamma^{-1}$ the mean life time, describes to great accuracy
the decay of unstable nuclei. It has found its theoretical derivation in
quantum mechanics by means of the Fermi Golden rule and it can be easily
obtained by assuming the Breit-Wigner distribution for the energy of the
unstable state. However, such distribution, while very useful in describing
data, is based on the assumption that $\Gamma$ is very small in comparison
with the mass/energy of the state and, more important, suffers the problem of
the absence of a minimum value for the energy, which in turn would correspond
to an Hamiltonian unbounded from below.

It is actually a renowned fact that in quantum mechanics $p(t)$ does not
exactly follow an exponential law. Deviations in the short- and long-time
regimes take place, see the general discussion in
Ref.~\cite{1978RPPh...41..587F}. In particular, by writing $p(t)=e^{-\gamma
(t)t}$, the function $\gamma(t)$ is not constant but decreases for short times
and vanishes for $t\rightarrow0$ \cite{khalfin,2008JPhA...41W3001F}. This
property is at the origin of the so-called quantum Zeno effect
\cite{1977JMP....18..756M}, according to which subsequent measurements, and
therefore subsequent collapses of the wave function of the unstable state
$\left\vert S\right\rangle $ during the non-exponential regime, generate a
slower decay rate and, in the limit of a continuous observation, a complete
inhibition of the decay. In fact, after $N$ measurements performed at time
intervals $t_{\ast}$, the probability that the state has not decayed at the
time $T=Nt_{\ast}$ is given by $p(t_{\ast})^{N}=e^{-\gamma(t_{\ast})T},$ which
is larger than the survival probability obtained for a single measurement
performed at the instant $T,$ $p(T)=e^{-\gamma(T)T},$ as long as
$\gamma(t_{\ast})<\gamma(T)\simeq\Gamma.$ Moreover, $p(t_{\ast})^{N}%
\rightarrow1$ for $t_{\ast}\rightarrow0$ (i.e. for $N\rightarrow\infty$ by
keeping $T$ fixed) because $\lim_{t\rightarrow0^{+}}\gamma(t)=0 $, implying
that $S$ does not decay at all. Interestingly, there can be also values of
$t_{\ast}$ such that $\gamma(t_{\ast})>\Gamma$, thus the measurements would
originate a faster decay rate (anti-Zeno effect) \cite{2000Natur.405..546K}.
The quantum Zeno and anti-Zeno effects have indeed found experimental
confirmation in cold atoms experiments \cite{1990PhRvA..41.2295I}, a discovery
that renewed also the theoretical interest on these fascinating features of
quantum systems.

In the middle of the 20th century Relativistic Quantum Field Theory (RQFT) has
been developed. Since within this formalism the number of particles is not
conserved, RQFT has been recognized to be the most natural theoretical
framework for the study of decays. The fundamental randomness in the process
of creating and annihilating particles is at the root of the indetermination
of the lifetime of unstable particles. Electromagnetic decays of atoms can be
driven back to the emission of photons in the context of the best known RQFT,
Quantum Electrodynamics (QED). More in general, decay widths play nowadays an
important role in many phenomenological studies of the Standard Model (SM),
such as hadron decays in QCD and weak decays of leptons, heavy quarks and weak
bosons. Also the recent search at LHC for the last missing particle of the SM,
the Higgs boson, relies upon the predictions of its decay properties.

The evaluation of the decay width $\Gamma$ of an unstable state $\left\vert
S\right\rangle $ in RQFT is now a technically well-defined task. However, in
view of the present discussion, a natural question immediately emerges: Are
short-time deviations from the exponential law $p(t)=e^{-\Gamma t}$ present
also in the context of RQFT?

In the perturbative approach of Refs. \cite{1993PhRvL..71.2687B,Maiani:1997pd}
no (or very much suppressed) short-time deviations from the exponential law,
and thus no quantum Zeno effect, were found within RQFT. In this work, after a
critical reconsideration of the issue of the survival probability in RQFT, we
actually obtain the opposite answer: short-time deviations from the
exponential survival probability \emph{do} occur in a genuine RQFT context in
the case of superrenormalizable theories, to which we restrict our attention
here. In fact, $\gamma(t)<$ $\Gamma$ for short times and $\gamma
(t\rightarrow0^{+})=0$, implying that the quantum Zeno effect is possible also
in RQFT. More in general, our results are compatible with the non-relativistic
model of Ref. \cite{2002quant.ph..2127F}, in which a deeper understanding of
the quantum Zeno and anti-Zeno effects has been achieved. Remarkably, the
short-time deviation from the exponential law is strongly enhanced in the case
of the short-living hadrons, such as in the case of the decay $\rho
\rightarrow\pi\pi$ which will be explicitly investigated later. This is an
interesting result on its own, which might affect the dynamics of the fast
expanding hadrons fireball in heavy ions collisions experiments in which these
particles are abundantly produced.\textbf{\ }

We now turn to a concrete example by considering the following
(superrenormalizable) RQFT Lagrangian with two scalar fields $S$ and $\varphi
$:
\begin{equation}
\mathcal{L}=\frac{1}{2}(\partial_{\mu}S)^{2}-\frac{1}{2}M_{0}^{2}S^{2}%
+\frac{1}{2}(\partial_{\mu}\varphi)^{2}-\frac{1}{2}m^{2}\varphi^{2}%
+gS\varphi^{2}. \label{lag}%
\end{equation}
The interaction term $\mathcal{L}_{int}=gS\varphi^{2}$ induces the decay
process $S\rightarrow\varphi\varphi$, whose tree-level decay rate reads%
\begin{equation}
\Gamma_{S\varphi\varphi}^{\text{t-l}}=\frac{\sqrt{\frac{M_{0}^{2}}{4}-m^{2}}%
}{8\pi M_{0}^{2}}(\sqrt{2}g)^{2}\theta\left(  M_{0}-2m\right)  \text{ .}
\label{tl1}%
\end{equation}
The `naive', tree-level expression of the survival probability $p(t)$ for the
resonance $S$ created at $t=0$ is $p_{\text{t-l}}(t)=e^{-\Gamma_{S\varphi
\varphi}^{\text{t-l}}t}$ and the tree-level expression of the mean life time
is $\tau_{\text{t-l}}=1/\Gamma_{S\varphi\varphi}^{\text{t-l}}$.

A crucial intermediate step toward the determination of the survival
probability $p(t)$ within the RQFT framework is the evaluation of the
propagator $\Delta_{S}(p^{2})$ of the unstable resonance $S$, which is
obtained by (re)summing the one-particle irreducible self-energy contribution
$\Sigma(p^{2})$:%
\begin{equation}
\Delta_{S}(p^{2})=\left[  p^{2}-M_{0}^{2}+(\sqrt{2}g)^{2}\Sigma(p^{2}%
)+i\varepsilon\text{ }\right]  ^{-1}\text{.}%
\end{equation}
To lowest order, $\Sigma(p^{2})$ corresponds to a bubble of two fields
$\varphi$:
\begin{equation}
\Sigma(p^{2})=\int_{q}\frac{-i}{\left[  \left(  \frac{p+2q}{2}\right)
^{2}-m^{2}+i\varepsilon\right]  \left[  \left(  \frac{p-2q}{2}\right)
^{2}-m^{2}+i\varepsilon\right]  }\text{ ,} \label{loop}%
\end{equation}
with $\int_{q}=\int\frac{d^{4}q}{(2\pi)^{4}}.$ The spectral function
$d_{S}(x)$ of the scalar field $S$ is proportional to the imaginary part of
the propagator:%
\begin{equation}
d_{S}(x=\sqrt{p^{2}})=\frac{2x}{\pi}\left\vert \lim_{\varepsilon\rightarrow
0}\mathrm{Im}[\Delta_{S}(p^{2})]\right\vert \text{ .}%
\end{equation}
The quantity $d_{S}(x)dx$ represents the probability that in the rest frame of
$S$ the state $S$ has a mass between $x$ and $x+dx.$ It is correctly
normalized for each $g$, $\int_{0}^{\infty}d_{S}(x)dx=1$ and, in the limit
$g\rightarrow0,$ the expected spectral function $d_{S}(x)=\delta(x-M_{0})$ is
obtained \cite{Achasov:2004uq,Giacosa:2007bn}. This fact allows to determine
the probability amplitude $a(t),$ and therefore the survival probability
$p(t)$:%
\begin{equation}
a(t)=\int_{-\infty}^{+\infty}\mathrm{dx}\,\,d_{S}(x)e^{-ixt}\text{ ,
}p(t)=\left\vert a(t)\right\vert ^{2}\text{ .} \label{p(t)}%
\end{equation}
The condition $p(0)=1$ is fulfilled in virtue of the normalization of
$d_{S}(x)$. This property is, in turn, a consequence of the 1-loop resummation
and the validity of the K\"{a}llen-Lehman representation. (The integral in Eq.
(\ref{p(t)}) is actually limited to the interval $(2m,\infty)$ in virtue of
the step-function $\theta(x-2m)$ arising in $d_{S}(x),$ see the optical
theorem below. The extension to the integration range $(-\infty,\infty)$
allows to express $a(t)$ as the Fourier-transform of $d_{S}(x),$ which
represents a technical help in a variety of applications). The general
discussion on $p(t)$ in Refs.~\cite{1978RPPh...41..587F,2001PhRvL..86.2699F}
in the framework of quantum mechanics is applicable in the present RQFT theory.

An important aspect concerning the definition of the properties of unstable
states has been raised in Ref. \cite{Gegelia:2009py} where it has been pointed
out that physical measurable quantities must be invariant under field
redefinitions, as the $S$-matrix elements, leading to the so called
\textquotedblleft complex mass renormalization scheme\textquotedblright. A
natural question arises whether the survival probability of Eq.~(\ref{p(t)})
is invariant under field redefinition. In Appendix A the issue is discussed in
more detail: a redefinition of the fields corresponds to a change of the
initial state $\left\vert S\right\rangle $. A related subtle point concerns
the state formation at $t=0$: in Ref.~\cite{Maiani:1997pd} the full scattering
process $\varphi\varphi\rightarrow S\rightarrow\varphi\varphi$ is computed to
second order in perturbation theory and also the \textquotedblleft formation
time\textquotedblright\ of the resonance has been modeled. Even if there is no
instant of time at which the state of the system corresponds to the state $S$,
the survival amplitude $a(t)$ directly enters in the calculation of the
temporal evolution of the system and could therefore lead to \textquotedblleft
observable\textquotedblright\ effects (see Appendix B for a detailed discussion).

For what concerns the measurability of the spectral function $d_{S}(x),$ we
devise the following situation: we introduce two scalar fields $A$ and $B,$
the first massless and the second with mass $M_{B}>M_{S}$ and write down the
interaction Lagrangian $\mathcal{L}_{int}=cBAS$ . We suppose that the
interaction strength $c$ is small enough to allow for a tree-level analysis of
the process $B\rightarrow AS,$ which reads $\Gamma_{BAS}^{\text{t-l}}%
(M_{B})=\frac{p_{BAS}}{8\pi M_{B}^{2}}c^{2}.$ When $g\neq0$ the state $S$
decays into $\varphi\varphi$. Physically, one observes a tree-body decay
$B\rightarrow A\varphi\varphi,$ whose decay-rate reads:%
\begin{equation}
\Gamma_{BA\varphi\varphi}^{\text{t-l}}(M_{B})=\int_{0}^{M_{B}}\Gamma
_{BAS}^{\text{t-l}}(M_{B})d_{S}(x)dx. \label{3bd}%
\end{equation}
The tree-body decay is decomposed into two steps: $B\rightarrow AS$ and
$S\rightarrow\varphi\varphi.$ The quantity $\Gamma_{BAS}^{\text{t-l}}(M_{B})$
represents the decay rate for $B\rightarrow AS$ (at a given mass $x $ for the
state $S$) and $d_{S}(x)dx$ is the corresponding weight, i.e. the probability
that the resonance $S$ has a mass between $x$ and $x+dx.$ In this simple
example the spectral function $d_{S}(x)$ emerges naturally as a mass
distribution, correctly normalized, for the scalar state $S.$ By measuring the
line shape of the particle $B$ (via the decay products $\varphi\varphi$ and
$A$) it is then possible to measure the quantity $\Gamma_{BAS}^{\text{t-l}%
}(M_{B})d_{S}(x),$ and therefore the mass distribution $d_{S}(x).$ This
example shows that, within the framework of the introduced toy Lagrangians,
the quantity $d_{S}(x)$ can be `measured'. Interestingly, there are
experimental situations which are conceptually similar to the here presented
case: the decay $\phi\rightarrow\gamma\pi^{0}\pi^{0}$ through the intermediate
$a_{0}(980)$ and $f_{0}(980)$ mesons \cite{Giacosa:2008st}, the similar decay
of the $j/\psi$ charmonium \cite{Bugg:2006sr}, or the hadronic decay of the
$\tau$ lepton into $\nu\pi\pi,$ dominated by the $\rho$ meson for an invariant
$\pi\pi$ mass close to $\rho$ mass \cite{Schael:2005am}. It should be clearly
stressed that the mentioned experiments are by far not so clean as our
depicted toy model due to the presence of many possible intermediate states
and background interactions. Moreover, the exact theory of hadrons, being not
derivable from QCD, is unknown and therefore the determination of hadronic
spectral functions is in most cases model dependent. Our attention to hadrons,
specifically to the example of the $\rho$ meson later on, is thus limited to
simple hadronic models. However, here we are not interested to a precision
study of hadronic spectral functions, but only to the order of magnitude
involved in the deviation from the exponential decay law, for which a
simplified treatment of hadrons is --at least as a first step-- justifiable.

After this digression on the spectral functions, we turn to the main subject
of this work, which is the behavior of the survival probability $p(t).$ The
first derivative of $p(t)$ is well defined and vanishes, $p^{\prime}(t=0)=0$
as a consequence of the fact that the integral $\int_{0}^{\infty}%
x\,d_{S}(x)dx$ is finite and real (it is the mean mass $\left\langle
M\right\rangle $, a reasonable definition for the mass of a resonance
\cite{Giacosa:2007bn}). This, in turn, implies that the function
$\gamma(t)=\frac{-1}{t}\ln p(t)$ vanishes for $t\rightarrow0^{+}$:%
\begin{equation}
\lim_{t\rightarrow0^{+}}\gamma(t)=-\lim_{t\rightarrow0^{+}}\frac{p^{\prime
}(t)}{p(t)}=0.
\end{equation}
We can therefore conclude that the quantum Zeno effect is perfectly possible
in the present RQFT context.

In order to explicitly calculate the function $p(t)$ one has first to evaluate
the loop integral. In the rest frame of the $S$ particle ($p=(x,0)$) one first
solves the integral over $q^{0}$ by calculating the residues and then
introduces a cutoff $\Lambda$ on the remaining integral over $d^{3}q$
obtaining:
\begin{align}
\Sigma(x)  &  =\frac{-\sqrt{4m^{2}-x^{2}}}{8\pi^{2}x}\arctan\left(
\frac{\Lambda x}{\sqrt{\Lambda^{2}+x^{2}}\sqrt{4m^{2}-x^{2}}}\right)
\nonumber\\
&  -\frac{1}{8\pi^{2}}\log\left(  \frac{m}{\Lambda+\sqrt{\Lambda^{2}+m^{2}}%
}\right)  .
\end{align}
A general property for $\Sigma(x)$ follows from the optical theorem:
\[
I(x)=(\sqrt{2}g)^{2}\mathrm{Im}[\Sigma(x)]=x\Gamma_{S\varphi\varphi
}^{\text{t-l}}(x)\theta\left(  \sqrt{\Lambda^{2}+m^{2}}-\tfrac{x}{2}\right)
\text{.}%
\]
The function $I(x)$ is zero for $0<x<2m$ and for $x>2\sqrt{\Lambda^{2}+m^{2}}$
and -in between- does not depend on the cutoff $\Lambda$. The quantity
$R(x)=(\sqrt{2}g)^{2}\mathrm{Re}[\Sigma(x)]$ is nonzero below and above
threshold and depends explicitly on $\Lambda$. The physical (Breit-Wigner)
mass $M$ of the scalar field $S$ is modified by the 1-loop corrections and is
determined by the equation: $M^{2}-M_{0}^{2}+R(M)=0$ In general, $M\neq
M_{0}.$ However, the requirement $M=$ $M_{0}$ can be fulfilled by introducing
a counterterm in Eq. (\ref{lag}): $\mathcal{L\rightarrow L-}\frac{1}{2}CS^{2}$
with $C=R(M_{0})$. (Note, one could well work with a physical mass $M\neq
M_{0},$ provided that the tree-level decay width in Eq. (\ref{tl1}) is
evaluated at the physical mass $M$.)

There are basically two different approaches to deal with the described set of
equations: (i) the theory is regarded as a fundamental theory valid up to
-say- the Planck energy; (ii) the theory is regarded as an effective,
low-energy manifestation of some other theory and the cutoff $\Lambda$ is a
finite number of the same order of magnitude of the masses. In the following
we study separately these two cases.

\emph{Case (i): }$\mathcal{L}$ \emph{as `fundamental' theory}: When the cutoff
$\Lambda$ is much larger than the other scales of the model, as in the case
$\Lambda\simeq M_{Planck},$ it is convenient and numerically exhaustive to
perform the limit $\Lambda\rightarrow\infty.$ In order to have a finite
physical mass $M=M_{0},$ the counterterm $C=R(M_{0})$ needs to be very large
(formally divergent, $\Lambda\rightarrow\infty$). Once this divergence has
been subtracted, all the results -including the survival probability $p(t)$-
are finite and well defined.

\begin{figure}[ptb]
\begin{centering}
\epsfig{file=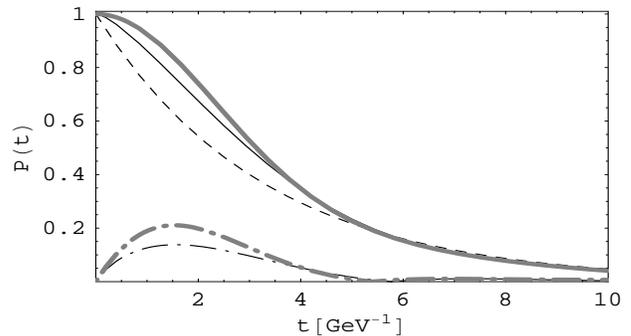,height=4.5cm,width=8.5cm,angle=0}
\caption{The survival probability $p(t)$ of Eq. (\ref{p(t)}) is shown in the
case of infinite (thin solid line) and finite, $\Lambda=1$ GeV, (thick gray line) cutoff. In
both cases the non-exponential behavior at short times is clearly visible. The
exponential tree-level decay is shown for comparison (dashed line).
The quantity $\left\vert p(t)-e^{-\Gamma t} \right\vert$
is also displayed by the thin dot-dashed and thick gray dot-dashed lines for the
two cases respectively.}
\end{centering}
\end{figure}

Now we turn to a quantitative estimate of the short-time interval in which the
deviations from the exponential decay law are non-negligible. In the
Literature a Taylor expansion of the function $p(t)$ is often performed and
the so-called Zeno time $\tau_{Z}=\sqrt{-2/p^{^{\prime\prime}}(0)}$ is
introduced as a measure of the short-time interval with a non-exponential
behavior. This procedure is, however, not general: it is in fact a priori not
obvious that the second derivative $p^{\prime\prime}(0)$ exists. In the
present case, for instance, the latter diverges for $\Lambda\rightarrow
+\infty$ since $d_{S}(x)$ behaves asymptotically as $1/x^{3}$. We thus
introduce a more general definition, which does not depend on the higher
derivatives of $p(t)$ at $t=0$. The time $\tau_{M}$ is defined as the instant
of time at which the deviation of the function $p(t)$ from the exponential
behavior $e^{-\Gamma t}$ is maximal:%
\begin{equation}
\max\left(  p(t)-e^{-\Gamma t}\right)  \rightarrow t=\tau_{M}.
\end{equation}
Clearly, $\frac{d}{dt}\left(  p(t)-e^{-\Gamma t}\right)  _{t=\tau_{M}}=0$. In
all practical cases $\tau_{M}$ corresponds to the first root of the derivative
of the function $p(t)-e^{-\Gamma t}.$

\begin{figure}[ptb]
\begin{centering}
\epsfig{file=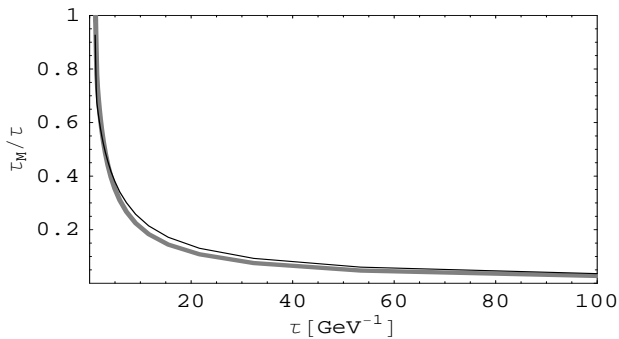,height=4.50cm,width=8.5cm,angle=0}
\caption{$\tau_{M}/\tau_{\text{t-l}}$ as function of $\tau_{\text{t-l}}$ in
the cases of infinite (thin line) and finite ($\Lambda=1,$ thick gray line)
cutoffs. For short living particles, such as hadronic resonances, the non exponential regime lasts for a time scale of the same order of magnitude of the mean life time of the particle.}
\end{centering}
\end{figure}

We now turn to a numerical example. We choose the physical mass as $M=M_{0}=1
$ GeV and $m=m_{\pi}=0.139$ GeV (typical values of hadronic particles). In
Fig.~1 the function $p(t)$ is plotted (solid thin line) for the choice
$g=2\sqrt{2}$ GeV$,$ which corresponds to a tree-level decay width
$\Gamma_{S\varphi\varphi}^{\text{t-l}}=$ $305.7$ MeV (on the high side of a
typical hadronic decay) and to a tree-level lifetime $\tau_{\text{t-l}}=3.27 $
GeV$^{-1}.$ The existence of a non-exponential behavior is clearly visible;
numerically, one obtains $\tau_{M}/\tau_{\text{t-l}}=0.48$, implying that the
non-exponential regime lasts an amount of time comparable with the mean life
time. Also the function $\left\vert p(t)-e^{-\Gamma t}\right\vert $ is
displayed (thin dot-dashed line) to clearly show the existence of $\tau_{M}$.

In Fig. 2 the ratio $\tau_{M}/\tau_{\text{t-l}}$ (thin solid line) is plotted
as a function of $\tau_{\text{t-l}}$ (i.e., as a function of the coupling
constant $g^{-2}$). The ratio $\tau_{M}/\tau_{\text{t-l}}$ increases for
decreasing $\tau_{\text{t-l}}$: the non-exponential regime is always present
but is enhanced for short living particles (lifetime typical of a strong
decay), while it decreases for long-living particles (i.e. in the regimes of
electromagnetic and weak decays). For instance, decreasing the coupling $g$ to
$1.15$ GeV implies a decay width of about $50$ MeV, which is on the low side
of a typical hadronic decay (as, for instance, the meson $f_{0}(980)$). In
this case, $\tau_{\text{t-l}}\simeq20$ GeV$^{-1}$, corresponding to a ratio
$\tau_{M}/\tau_{\text{t-l}}\simeq0.16,$ which is still a sizable quantity. We
thus conclude that the non-exponential regime for a typical hadronic decays
amounts to $15$-$50\%$ of the mean lifetime.

\emph{Case (ii): }$\mathcal{L}$ as \emph{effective hadronic theory: }When the
toy model is interpreted as a prototype of an hadronic effective theory, the
cutoff is a further parameter entering in the model with a typical value of
about $\sim1$ GeV. (For definiteness we set $\Lambda=1$ GeV \cite{foot}). The
results are qualitatively similar to the case $\Lambda\rightarrow+\infty,$ but
the existence of a finite cutoff increases the size of the short-time
deviations from the exponential law, as clearly visible in Fig.~1 for
$g=2\sqrt{2}$ GeV (thick gray line). Indeed in this case the second derivative
of $p(t)$ is finite at $t=0$ and the usual quadratic approximation for $p(t)$
at short times could be adopted. In Fig.~2 the ratio $\tau_{M}/\tau
_{\text{t-l}}$ is shown as a function of $\tau_{\text{t-l}}$ for $\Lambda=1$
GeV and is quite similar to the previous case. However, while the value of
$\tau_{M}$ is almost independent from the choice of the cutoff, the difference
between $p(t)$ and the exponential decay law is instead larger in the case of
a finite cutoff. \begin{figure}[ptb]
\begin{centering}
\epsfig{file=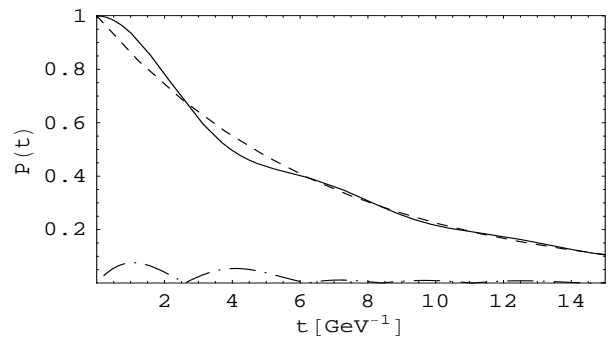,height=4.5cm,width=8.5cm,angle=0}
\caption{Survival probability $p(t)$ (solid line) in the case of the $%
\protect\rho$ meson and corresponding exponential decay law. The non exponential behavior (quadratic in this case)
is clearly visible at short times. The crossing of the $p(t)$ with the exponential decay indicate the possible occurrence
of also the Anti-Zeno effect. The quantity $\left\vert p(t)-e^{-\Gamma t} \right\vert$ is also shown (dot-dashed line).}
\end{centering}
\end{figure}

Bearing in mind all the previously mentioned caveats of hadronic spectral
functions, it is anyhow interesting to conclude the present study with a
physical example. To this purpose scalar states are not suitable because their
masses and decay widths are often affected by large errors, see
\cite{Yao:2006px}. We consider instead the $\rho$ meson, whose mass and (by
far dominant) decay into two pions are very well measured: $M_{\rho}=775\pm1$
MeV, $\Gamma_{\rho\rightarrow\pi\pi}=149\pm0.5$ MeV ~\cite{Yao:2006px}. The
mass distribution reads:
\begin{equation}
d_{\rho}(x)=\frac{2x}{\pi}\frac{x\Gamma_{\rho\rightarrow\pi\pi}(x)}%
{(x^{2}-M_{\rho}^{2})^{2}+x^{2}\Gamma_{\rho\rightarrow\pi\pi}(x)^{2}}\text{ ,
} \label{drho}%
\end{equation}
where $\Gamma_{\rho\rightarrow\pi\pi}(x)=\frac{\left(  \frac{x^{2}}{4}%
-m^{2}\right)  ^{3/2}}{6\pi x^{2}}g_{\rho}^{2}$ and $g_{\rho}=5.98.$
The ratio $\tau_{M}/\tau_{\text{t-l}}=0.16$ implies that, also in this
concrete case, a sizable interval of non-exponential regime holds. Note, the
function $p(t)$ crosses the exponential function $e^{-\Gamma_{\rho
\rightarrow\pi\pi}t},$ thus indicating the existence also of an anti-Zeno
regime \cite{2001PhRvL..86.2699F}. It will be interesting to study to which
extent our results affect the evolution of the hot and expanding gas of
hadrons produced in heavy ions collisions experiments and the spectra of the
particles emitted by the plasma. In the presently available transport
simulations indeed a simple exponential decay law is assumed for the hadronic
resonances \cite{Peters:1997va} whereas during the very short characteristic
time scale of the evolution of the plasma sizable deviations from the
exponential law are present.

The present work is based on the (resummed) 1-loop approximation. Future
studies should go beyond this scheme and include higher order terms, the first
one being the `sunset' diagram, in which a particle $S$ is exchanged by the
two particles $\varphi$ circulating in the loop. The non-exponential nature of
the decay does not depend on the truncation and would take place also when of
higher order contributions are included. The numerical influence of the latter
is, however, not expected to be large: higher order amplitudes are suppressed
in the so-called large-$N_{c}$ approximation \cite{witten}, which is a
successful scheme for hadronic theories. (The sunset diagram is suppressed of
$1/N_{c}$ w.r.t. the calculated loop). Moreover, the higher the order, the
larger the number of vertex functions, which suppress the corresponding
self-energy amplitude. It is then expected that higher order terms do not
change the picture presented in this work, although the explicit verification
of this statement is left as an interesting outlook. Two further natural
outlooks of the present work are: (i) The study of short-time deviations from
the exponential law in the context of renormalizable and non-renormalizable
RQFT Lagrangians. (ii) The study of an unstable resonance decaying in two (or
more) channels. Both aim to a deeper understanding of decay processes in RQFT
and can find various applications in the context of strongly decaying hadrons
and the other decay processes of the Standard Model.

\bigskip

The work of G.~P. is supported by the Deutsche Forschungsgemeinschaft (DFG)
under Grant No. PA 1780/2-1. We also thank E.~Santini for valuable discussions.

\section{Appendix A}

Field redefinitions do not change the physical content of the theory, as e.g.
the $S$-matrix elements for asymptotic initial and final states
\cite{Kamefuchi:1961sb}. However, the Green-functions, and in particular the
propagator of an unstable particle, are not invariant under field
redefinitions, see Ref. \cite{Gegelia:2009py} and refs. therein. It is then
important to study the influence of field redefinitions on the results of the
present work. For definiteness we consider the following transformation with
unit Jacobian:
\begin{equation}
S\rightarrow\tilde{S}=S-\alpha\varphi^{2}\text{ , }\varphi\rightarrow
\tilde{\varphi}=\varphi\text{ ,} \label{trasform}%
\end{equation}
where $\alpha$ is a dimensionful constant. In this way we go from the
representation 1 (in terms of the fields $\{S,\varphi\},$ whose Lagrangian
$\mathcal{L}_{repr1}(S,\varphi)=\mathcal{L}$ is given by Eq. (\ref{lag})) to
the representation 2 (in terms of the fields $\{\tilde{S},\tilde{\varphi}\}$).
In the representation 2 the Lagrangian $\mathcal{L}_{repr2}(\tilde{S}%
,\tilde{\varphi})$ reads%

\begin{equation}
\mathcal{L}_{repr2}(\tilde{S},\tilde{\varphi})=\mathcal{L}_{repr1}(S=\tilde
{S}+\alpha\varphi^{2},\varphi=\tilde{\varphi})\text{ .}%
\end{equation}
Also in term of the Hamiltonians, $H_{repr1}(S,\varphi)$ in representation 1
and $H_{repr2}(\tilde{S},\tilde{\varphi})$ in representation 2 one has
\begin{equation}
H_{repr2}(\tilde{S},\tilde{\varphi})=H_{repr1}(S=\tilde{S}+\alpha\varphi
^{2},\varphi=\tilde{\varphi}). \label{ham}%
\end{equation}
$H_{repr1}$ is written as $H_{repr1}=H_{0,repr1}+H_{1,repr1}$ where as usual
the `non interacting part' $H_{0,repr1}$ is given by
\begin{align}
H_{0,repr1}  &  =\int d^{3}x\frac{1}{2}\left(  \partial_{0}S\right)
^{2}+\frac{1}{2}\left(  \bigtriangledown S\right)  ^{2}+\frac{M_{0}^{2}}%
{2}S^{2}\nonumber\\
&  +\frac{1}{2}\left(  \partial_{0}\varphi\right)  ^{2}+\frac{1}{2}\left(
\bigtriangledown\varphi\right)  ^{2}+\frac{m^{2}}{2}\varphi^{2}.
\end{align}
Similarly, $H_{repr2}$ is written as $H_{repr2}=H_{0,repr2}+H_{1,repr2}$ where
by definition:
\begin{equation}
H_{0,repr2}(\tilde{S},\tilde{\varphi})=H_{0,repr1}(\tilde{S},\tilde{\varphi
})\text{ .}%
\end{equation}
This implies that $H_{0,repr1}$ and $H_{0,repr2}$ have the same functional
form. However, it is important to stress that the two operators are different:
$H_{0,repr2}(\tilde{S},\tilde{\varphi})\neq H_{0,repr1}(S,\varphi)$. This can
be easily proven by plugging Eqs.~(\ref{trasform}) into $H_{0,repr2}(\tilde
{S},\tilde{\varphi})$: one obtains the operator $H_{0,repr2}(\tilde{S}%
,\tilde{\varphi})=H_{0,repr2}(S-\alpha\varphi^{2},\varphi)$ which is indeed
-in terms of $S$ and $\varphi$- a complicated Hamiltonian.

In this work we have calculated the survival probability
\begin{equation}
a(t)=\left\langle S\right\vert e^{-iH_{repr1}(S,\varphi)t}\left\vert
S\right\rangle \text{ }%
\end{equation}
where $\left\vert S\right\rangle $ is an eigenstate of $H_{0,repr1}$ with
eigenvalue $M_{0}$ (and with three-momentum $\vec{P}=0$). This quantity is
indeed, in virtue of Eq. (\ref{ham}), invariant under the choice of representation.

However, if one would repeat the calculation of the survival probability in
representation $2$ using the same mathematical approach leading to Eq.
(\ref{p(t)}), one would calculate the quantity
\begin{equation}
\tilde{a}(t)=\left\langle \tilde{S}\right\vert e^{-iH_{repr2}(\tilde{S}%
,\tilde{\varphi})t}\left\vert \tilde{S}\right\rangle \text{ ,}%
\end{equation}
where $\left\vert \tilde{S}\right\rangle $ is the eigenstate with energy
$M_{0}$ (and $\vec{P}=0$) of the operator $H_{0,repr2}(\tilde{S}%
,\tilde{\varphi})\neq H_{0,repr1}(S,\varphi)$. It should be stressed that
\begin{equation}
\left\vert \tilde{S}\right\rangle \neq\left\vert S\right\rangle .
\end{equation}
Naively, the state $\left\vert \tilde{S}\right\rangle \simeq\left\vert
S\right\rangle +\alpha\left\vert \varphi\varphi\right\rangle $ (with proper
normalizations and also including the proper regularizations) is a
superposition of the state $\left\vert S\right\rangle $ with the two-body
state $\left\vert \varphi\varphi\right\rangle $ (which includes a sum over
internal momenta, which we do not specify here). It is then clear that
$\tilde{a}(t)\neq a(t)$, but this is an effect of changing the initial state,
$\left\vert \tilde{S}\right\rangle \neq\left\vert S\right\rangle .$ This is
also the reason why the mass distributions $d_{S}(x)$ and $d_{\tilde{S}}(x)$
(which are the imaginary part of the propagators of $S$ and $\tilde{S}$ in the
first and second representations, respectively) do not coincide. In order to
be consistent and to calculate the same quantity in the second representation,
one should not start from the initial state $\left\vert \tilde{S}\right\rangle
,$ but from the state $\left\vert S\right\rangle \simeq\left\vert \tilde
{S}\right\rangle -\alpha\left\vert \varphi\varphi\right\rangle .$ In this way
one would obtain the quantity $a(t)$ also in representation 2. It is then
evident from the present discussion that the representation choice is
intimately connected with the definition of the initial state of the system at
$t=0.$ More about this is discussed in the next Appendix.

Notice that if we perform the field transformation (\ref{trasform}) on the toy
Lagrangian $\mathcal{L}_{int}=cBAS$ we obtain $\mathcal{L}_{int}=cBA\tilde
{S}+c\alpha BA\varphi^{2}$. Now, in the evaluation of the three-body decay
$B\rightarrow A\varphi\varphi$ there is not only the intermediate state
$\tilde{S}$ because the new interaction $BA\varphi^{2}$ has emerged. For this
reason the theoretical result for the line shape is not $\Gamma_{BAS}%
^{\text{t-l}}(M_{B})d_{\tilde{S}}(x).$ The new term $BA\varphi^{2}$ generates
an interference with the amplitude given from the exchange of $\tilde{S}$, in
such a way that the final result -in agreement with the equivalence theorem-
coincides with Eq. (\ref{3bd}). Thus, when speaking about the mass
distribution we should always be aware that the discussion is valid in a given
representation. A change of representation generates a change of the state $S$
and therefore also the propagator and its imaginary part are modified.

\section{Appendix B}

Since we are dealing with unstable and short living particles one should also
consider the mechanism by which these resonances are created. The most
complete framework is the scattering $\varphi\varphi\rightarrow S\rightarrow
\varphi\varphi$. A full treatment implies the consideration of the wave
packets with proper initial conditions leading to some non-negligible spatial
overlap at -say- the time $t=0$. In the framework of plane waves, the full
state of the system can be expressed in terms of the eigenstates of the
Hamiltonian $H_{0}$:
\begin{equation}
\left\vert s(t)\right\rangle =\sum_{\mathbf{k}}c_{\mathbf{k}}(t)\left\vert
\varphi_{\mathbf{k}}\varphi_{\mathbf{-k}}\right\rangle +c_{S}(t)\left\vert
S\right\rangle .\nonumber
\end{equation}
The coefficient $c_{S}(t)$ is practically zero for $t<<0$ and only for
$t\simeq0$ it becomes significant. The way in which this happens can be
possibly considered as the formation process. If it were possible to tune the
starting conditions in such a way that $c_{S}(0)=1,$ we would have $\left\vert
s(t=0)\right\rangle =\left\vert S\right\rangle .$ From this point on, the
evolution is obtained by applying the time evolution operator. However, in
general the state at $t=0$ is a superposition:
\begin{equation}
\left\vert s(0)\right\rangle =\sum_{\mathbf{k}}c_{\mathbf{k}}(0)\left\vert
\varphi_{\mathbf{k}}\varphi_{\mathbf{-k}}\right\rangle +c_{S}(0)\left\vert
S\right\rangle .\nonumber
\end{equation}
Further evolution implies:
\begin{align}
&  e^{-iHt}\left\vert s(0)\right\rangle =\nonumber\\
&  \sum_{\mathbf{k}}c_{\mathbf{k}}(0)e^{-iHt}\left\vert \varphi_{\mathbf{k}%
}\varphi_{\mathbf{-k}} \right\rangle +c_{S}(0)e^{-iHt}\left\vert
S\right\rangle =\nonumber\\
&  \sum_{\mathbf{k}}c_{\mathbf{k}}(0)e^{-iHt}\left\vert \varphi_{\mathbf{k}%
}\varphi_{\mathbf{-k}} \right\rangle +c_{S}(0)\left(  a(t)\left\vert
S\right\rangle +\left\vert \varphi\varphi\right\rangle \right)  .\nonumber
\end{align}
Clearly, the amplitude $a(t)$ is part of a more general expression. The
situation is of course more complicated, because we cannot evaluate properly
the quantity $e^{-iHt}\left\vert \varphi_{\mathbf{k}}\varphi_{\mathbf{-k}%
}\right\rangle $. It is indeed interesting to observe that, if $e^{-iHt}%
\left\vert \varphi_{\mathbf{k}}\varphi_{\mathbf{-k}}\right\rangle $ does not
contain the state $\left\vert S\right\rangle $ (for instance, if the two wave
packets are already far apart at $t>0$), then (up to a phase):
$a(t)=e^{-\Gamma t/2}$, i.e. the exponential regime is realized. As also
discussed in Ref.~\cite{1978RPPh...41..587F}, the rescattering processes,
which can occur if the two wave packets are close to each other, are
responsible for the non-exponential behavior.


\end{document}